\documentclass[aps,prl,twocolumn,showpacs,superscriptaddress,groupedaddress]{revtex4}  
\usepackage{graphicx}  
\usepackage{dcolumn}   
\usepackage{bm}        
\usepackage{amssymb}   
\usepackage{placeins}
\usepackage{amsmath}

\usepackage[paperwidth=216mm,paperheight=280mm,centering,hmargin=2cm,vmargin=2.5cm]{geometry}



\begin{document}
\author{A.J.~Goodman}
\affiliation{Department of Chemistry, Massachusetts Institute of Technology}
\author{W.A.~Tisdale}
\affiliation{Department of Chemical Engineering, Massachusetts Institute of Technology}
\email{tisdale@mit.edu}

\widetext
\leftline{Version 4 as of \today}
\leftline{Primary authors: Aaron G.}
\leftline{To be submitted to (PRL)}


\title{Enhancement of Second-Order Nonlinear Optical Signals by Optical Stimulation}
\date{\today}


\begin{abstract}
Second-order nonlinear optical interactions such as sum- and difference-frequency generation are widely used for bioimaging and as selective probes of interfacial environments. However, inefficient nonlinear optical conversion often leads to poor signal-to-noise ratio and long signal acquisition times. Here, we demonstrate the dramatic enhancement of weak second-order nonlinear optical signals via stimulated sum- and difference-frequency generation. We present a conceptual framework to quantitatively describe the interaction and show that the process is highly sensitive to the relative optical phase of the stimulating field. To emphasize the utility of the technique, we demonstrate stimulated enhancement of second harmonic generation (SHG) from bovine collagen-I fibrils. Using a stimulating pulse fluence of only 3 nJ/cm$^2$, we obtain an SHG enhancement $>$$10^4$ relative to the spontaneous signal. The stimulation enhancement is greatest in situations where spontaneous signals are the weakest - such as low laser power, small sample volume, and weak nonlinear susceptibility - emphasizing the potential for this technique to improve signal-to-noise ratios in biological imaging and interfacial spectroscopy.
\end{abstract}

\pacs{42}
\maketitle


Second-order nonlinear optical interactions are useful for probing the electronic and vibrational properties of surfaces and interfaces, measuring time-resolved interfacial dynamics, and studying the structure of biological tissue. Because second-order nonlinear optical processes are dipole-forbidden in centrosymmetric media, second-order nonlinear signals are inherently surface and interface-selective. For instance, vibrational sum frequency generation (SFG) can inform our understanding of chemical bonding at solid surfaces and aqueous interfaces \cite{SFG_Spec_polymer,SFG_Spec_AqueousEnv_ChemRev2002} and time-resolved second harmonic generation (SHG) can be used to study the ultrafast dynamics of charge transfer at donor-acceptor interfaces \cite{Tisdale2010_SciencePaper,bad_SHG_Data,Visualize_ChargeSep_OrgBulkHet_ncomm}. In biological tissues, the inherent nonlinearity of SHG enables label-free 3D imaging of protein scaffolds \cite{SHG_microscopy_natProtoc2012}.

Second-order nonlinear optical experiments are often limited by low nonlinear conversion efficiencies. The efficiency of these nonlinear optical processes is determined by the nonlinearity of the sample, the volume of material probed by the laser beam, and the incident pulses' energies and durations.  Even with the arrival of ultrafast pulsed lasers, weakly nonlinear media do not support efficient non-resonant SHG and SFG. Increasing the incident laser fluence can increase the conversion efficiency, but is often accompanied by sample photodamage. For many experiments, signal photon count rates are $<$100-1000 Hz; in such scenarios, the experiment's signal-to-noise ratio (SNR) is bounded by $\sqrt{n}$ due to the unfavorable statistics of counting small numbers of photons, n.

Optical stimulation is an approach that has been successfully used to enhance other weak, inelastic scattering phenomena such as Raman and resonant inelastic X-ray scattering \cite{StimRaman_First,StimXray}. In order to increase the efficiency of these inelastic scattering processes, light of the scattered signal frequency is coincident on the material with the pump, seeding the nonlinear conversion of the pump to the signal frequency. In the case of Raman scattering, optical stimulation has enabled many new technologies such as spectrally tailored microscopy and label-free video-rate imaging \cite{SRS_Science2008,SRS_VideoRate_Science2010,STE_SRS}. While optical parametric amplification is an example of a stimulated second-order process used in many labs to shift the frequency of ultrafast laser pulses, the use of optical stimulation to enhance weak signals in SHG spectroscopy or imaging has not been demonstrated.

Here, we show the stimulated enhancement of SHG and difference frequency generation (DFG) in a configuration that is suitable for a wide variety of samples. We quantitatively describe the observed power and phase dependences using a coupled-wave formalism and achieve signal amplification of $>$$10^4$ in the biologically relevant sample collagen I. Our analyses indicate that the degree of signal amplification scales inversely with the sample's nonlinear susceptibility $\chi^{\left(2\right)}$ and the distance over which the stimulating and fundamental fields interact, signifying that optical stimulation is most advantageous in systems with the weakest spontaneous signals.

In order to realize stimulated SHG, it is necessary to overlap pulses at the fundamental and second harmonic frequency in space, time, and direction at the sample.  To achieve this, we used the modified Mach-Zehnder interferometer diagrammed in Fig.~1(a).  The output of a 76 MHz repetition rate Ti:sapphire oscillator producing 100 fs pulses in the near-infrared $\left(\lambda=830\:\mathrm{nm}\right)$ was split into two equal intensity beams. The stimulating light $\left(\lambda=415\:\mathrm{ nm}\right)$ was generated in one arm of the interferometer using a phase-matched nonlinear optical crystal, $\beta$-barium borate (BBO). A mechanical delay stage and a piezo-mounted mirror were used to control the overall time delay and the relative optical phase, respectively, between the fundamental and stimulating laser pulses. The two beams were recombined using a dichroic mirror and focused collinearly onto the sample. The signal generated at the second harmonic frequency was sent either to a balanced photodiode (for detection of stimulated signals) or to a photomultiplier tube and gated photon counter (for detection of spontaneous signals and absolute-intensity calibration of the photodiode).

In contrast to spontaneous SHG, stimulated SHG signals are no longer background-free due to the presence of the incident stimulating field. Following analogous strategies used in stimulated Raman scattering \cite{SRS_Science2008}, we modulated the fundamental beam at 3 kHz, as illustrated in Fig.~1(b), to separate incident second harmonic intensity from that which was generated within the sample. The stimulated SHG signal was then demodulated by a phase-sensitive lock-in amplifier.

Initial studies were performed using the ideal nonlinear medium BBO as a model system. When the fundamental and stimulating pulses are not overlapped in time, there is a small $2\omega$ signal at the chopping frequency due to spontaneous SHG. When the time delay, $\Delta t$, approaches zero, the SHG signal is dramatically enhanced due to stimulation, as shown in Fig.~1(c).

\begin{figure}[t!]
\begin{center}
\includegraphics[width=.85\columnwidth]{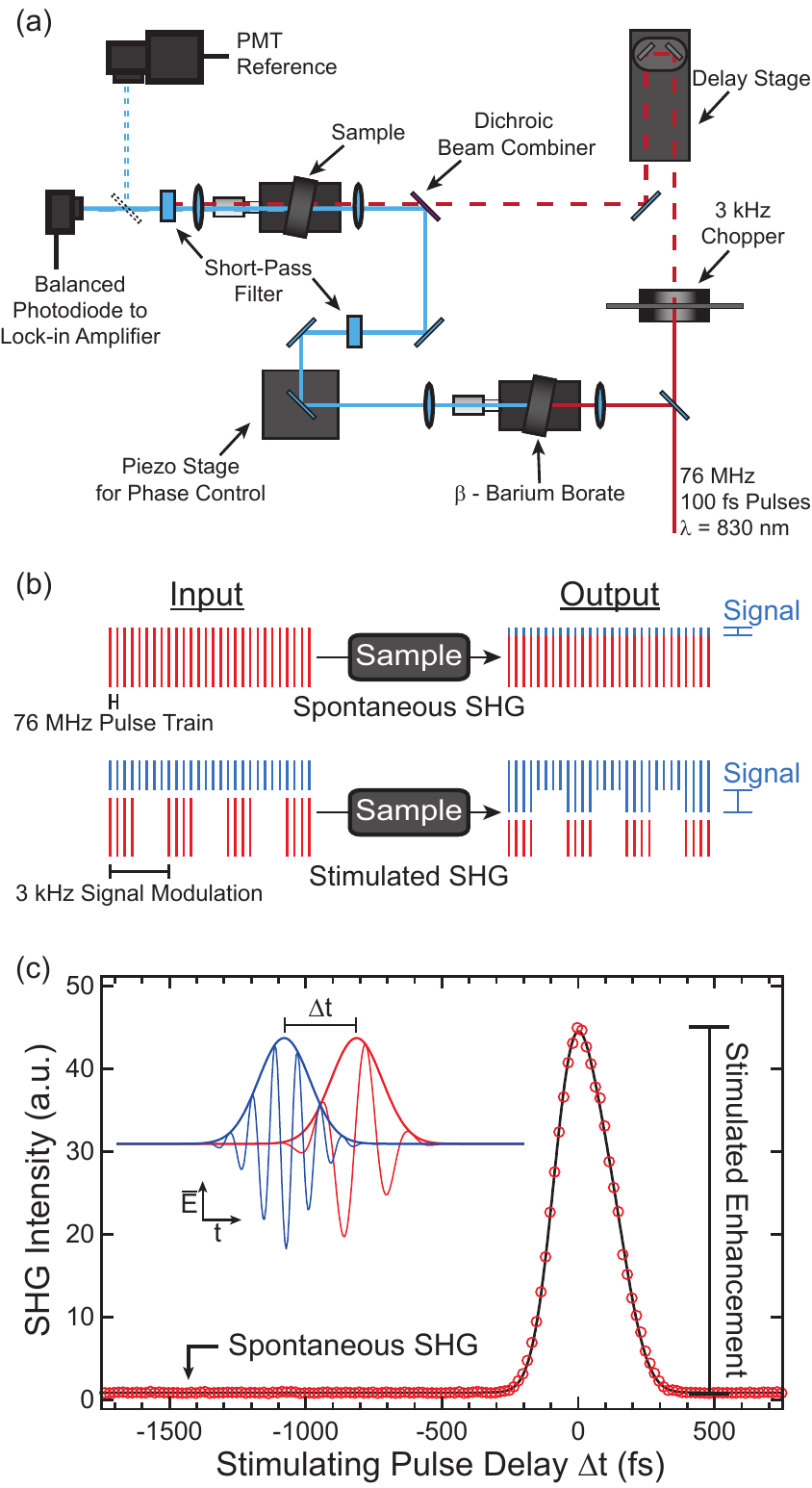}
\label{figure1}
\caption{(Color online) The stimulated SHG experiment (a) A schematic of the optical system used to perform stimulated SHG. (PMT:~photomultiplier~tube) (b) A visualization of the signal modulation scheme used to differentiate SHG signal from stimulating photons. (c) Stimulated SHG in $\beta-$barium borate.}
\end{center}
\end{figure}

To describe the interaction between the fundamental and stimulating fields we adapt some of the arguments made in the seminal 1962 paper by Armstrong $et$ $al.$ \cite{Armstrong1962}, with the assumption that there is no momentum mismatch between the fields.  When fields at the fundamental and second harmonic frequencies interact in a nonlinear medium, their amplitudes are coupled. The fields exchange energy over a characteristic length, $l$, where
\begin{align}
l^{-1}=2\omega^2\left(\frac{2\pi d_{\mathrm{eff}}}{c^2}\right)k_\omega^{-1}\sqrt{I_\mathrm{total}}\,.
\end{align}
\begin{figure*}[t!]
\centering
\includegraphics[width=\textwidth]{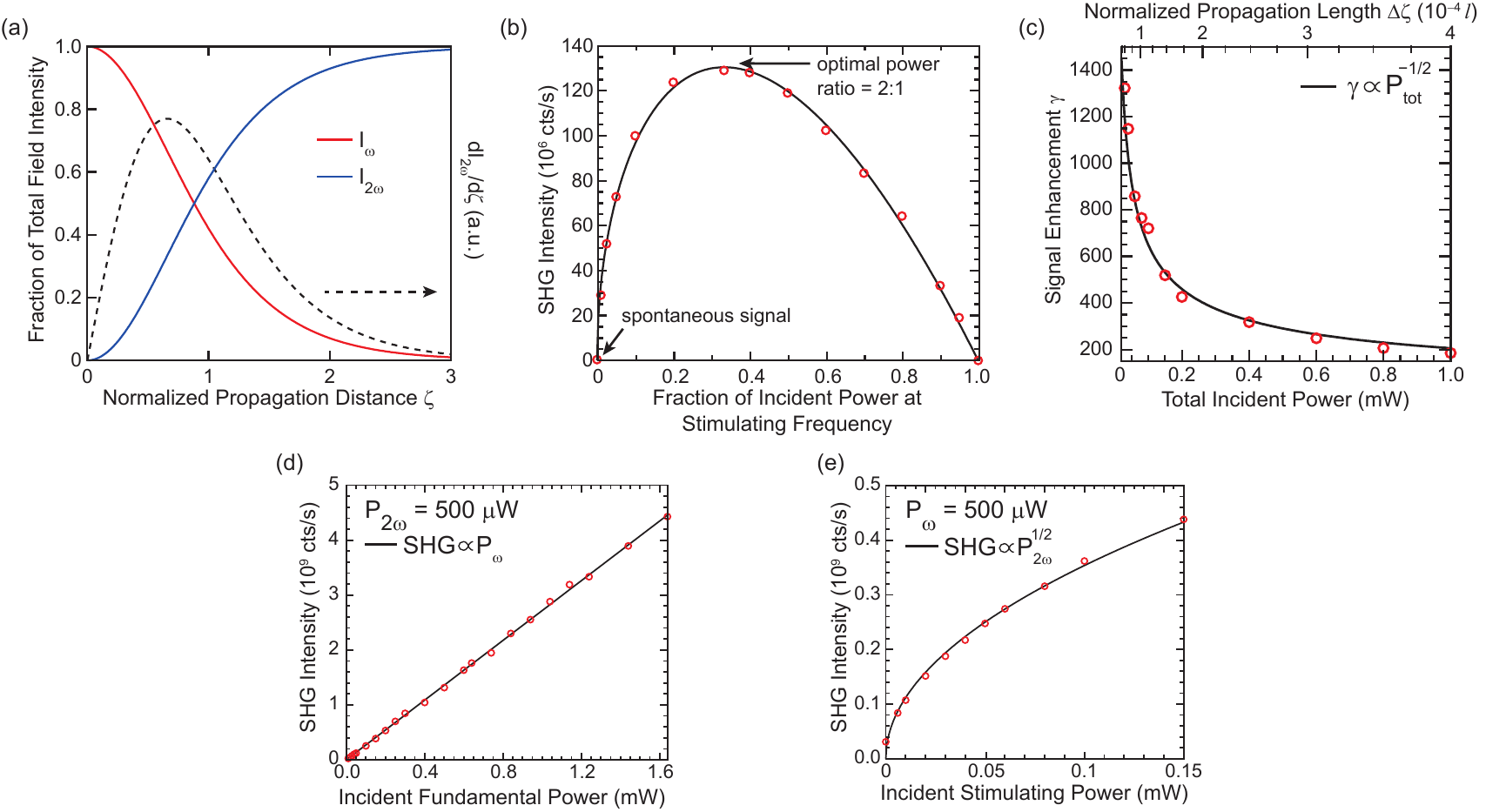}
\caption{(Color online) Analysis of stimulated SHG using the coupled-wave formalism.  Open circles are data from BBO, while black lines are analytical fits predicted by Eqs. (\ref{u1diff}$-$\ref{thetadiff}) (a) Solution of Eqs. (\ref{u1diff}$-$\ref{thetadiff}) with initial condition $\theta=-\frac{\pi}{2}$. (b) Dependence of the stimulated SHG signal on the composition of the incident fields. (c) Dependence of the signal enhancement $\gamma$ on the total incident intensity. (d) Dependence of the stimulated SHG signal on the fundamental intensity. (e) Dependence of the stimulated SHG signal on the stimulating intensity.}
\end{figure*}

\noindent Here, $d_{\mathrm{eff}}$ is the effective susceptibility, which takes into account the orientation of the fields with respect to the sample, $c$ is the speed of light, $k_\omega$ is the fundamental field wave vector, and $I_{\mathrm{total}}$ is the sum of the fundamental and second harmonic fields' intensities. If we normalize the field amplitudes to the total intensity, defining normalized field amplitudes $u_\omega$ and $u_{2\omega}$ such that
\begin{align}
u_\omega^2+u_{2\omega}^2=1
\end{align}
and defining the relative phase between the fundamental and stimulating waves to be
\begin{align}
\theta=2\phi_\omega-\phi_{2\omega},\label{theta_def}
\end{align}
then exchange of energy between the normalized fields is described by the coupled set of differential equations
\begin{align}
\frac{du_\omega}{d\zeta}&=u_\omega u_{2\omega}\sin(\theta)\label{u1diff}\\
\frac{du_{2\omega}}{d\zeta}&=-u_\omega^2\sin(\theta)\label{u2diff}\\
\frac{d\theta}{d\zeta}&=\frac{\cos(\theta)}{\sin(\theta)}\frac{d}{d\zeta}\left(\ln(u_\omega^2u_{2\omega})\right)\label{thetadiff}
\end{align}
where $\zeta=z/l$ is the [dimensionless] normalized propagation distance.

If the relative phase between the two waves is initially $\theta=\pm\frac{\pi}{2}$, then the relative phase does not change with propagation. The well-known $\theta=-\frac{\pi}{2}$ solution to Eqs.~(\ref{u1diff}$-$\ref{thetadiff}), plotted in Fig.~2(a), can be used to conceptually understand stimulated SHG. The derivative of the second harmonic intensity $dI_{2\omega}/d\zeta$ represents the rate of growth of the second harmonic field intensity; it is included in the figure as a dashed line. Spontaneous SHG occurs in the limit of $I_{2\omega}\rightarrow 0$ (the left edge of the plot), where conversion is slow and initiated by vacuum fluctuations. As $I_{2\omega}$ grows, its presence further accelerates conversion from $\omega$ to $2\omega$. Conceptually, the effect of stimulation can be understood as moving from a regime where $dI_{2\omega}/d\zeta$ is small to a regime where it is much larger. We note that introduction of an additional field in Eqs.~(\ref{u1diff}$-$\ref{thetadiff}) accounting for the stimulating beam is unnecessary, since the stimulating field is identical to the spontaneously generated field, and the rate of 2w generation isn't explicitly dependent on the field amplitude at prior zeta.

This effect can be seen clearly in Fig.~2(b), where the intensity of the second harmonic generated within the sample is plotted against the fraction of the incident intensity in the stimulating field, while the total intensity was held constant.  Data points on this plot can be mapped to $dI_{2\omega}/d\zeta$ at different values of $\zeta$ in Fig.~2(a). For fixed total incident power, optimal nonlinear conversion occurs when one third of the incident intensity is in the stimulating field, reflecting the maximum in $dI_{2\omega}/d\zeta$ visible in Fig.~2(a).

We define the ratio of the stimulated signal intensity to the spontaneous signal intensity (under conditions of constant total intensity and $2$:$1$ $I_\omega$:$I_{2\omega}$ stimulating ratio) as the “signal enhancement”, $\gamma$. By examining the power dependences of the spontaneous and optimal stimulated signals, we predict the behaviour of $\gamma$:
\begin{equation}
\begin{aligned}
\label{prop}
I_{\mathrm{sig}}^{\mathrm{stim}}&\propto I_{\mathrm{total}}^{3/2}\times d_{\mathrm{eff}}\times z\\
I_{\mathrm{sig}}^{\mathrm{spon}}&\propto I_{\mathrm{total}}^2\times d_{\mathrm{eff}}^2\times z^2\\
\gamma=I_{\mathrm{sig}}^{\mathrm{stim}}/I_{\mathrm{sig}}^{\mathrm{spon}}&\propto I_{\mathrm{total}}^{-1/2}\times d_{\mathrm{eff}}^{-1}\times z^{-1}
\end{aligned}
\end{equation}

\noindent Eq.~(\ref{prop}) implies that the signal enhancement $\gamma$ grows without bound as the extent of the nonlinear interaction decreases (i.e. the normalized propagation distance $\Delta\zeta$ decreases). Consequently, the advantage of stimulated SHG is greatest in precisely the situations where it is most needed: in weakly nonlinear media, small interaction volumes, or in media that do not admit the use of large incident powers. To demonstrate this point, the $I_{\mathrm{total}}^{-1/2}$ dependence of $\gamma$ was tested in BBO. The results are in agreement with the prediction of Eq.~(\ref{prop}) and are shown in Fig.~2(c). The dependences of stimulated SHG on $I_\omega$ with fixed $I_{2\omega}$ and on $I_{2\omega}$ with fixed $I_\omega$ are shown in Fig.~2(d)~and~(e) respectively, also in agreement with the predictions of Eq.~(\ref{prop}).

The relative optical phase $\theta$ defined in Eq.~(\ref{theta_def}) determines the nature of the nonlinear conversion. The $\theta$-dependence of Eqs.~(\ref{u1diff}-\ref{thetadiff}) suggests the possibility of changing the direction of energy flow between the fundamental and second harmonic fields. The two special cases $\theta=\pm\frac{\pi}{2}$ are illustrated in Fig.~3(a). When $\theta=-\frac{\pi}{2}$, SHG occurs, whereas a phase of $\theta=\frac{\pi}{2}$ induces DFG. Though the stimulated SHG and DFG signals have similar magnitudes, they have the opposite lock-in signal phase $\Omega$ relative to the modulation of the fundamental beam, as illustrated in Fig.~3(b). A stimulated SHG signal is in-phase with the reference modulation, while a stimulated DFG signal is $180^\circ$ out-of-phase. The relative optical phase of the two fields $–$ and, accordingly, the direction of energy flow between the fundamental and second harmonic $–$ can be finely controlled using a piezo-mounted mirror, as shown in Fig.~3(c). We note that unwanted phase fluctuations due to air currents and optical table vibrations are the dominant noise source in our experiments and present an additional difficulty compared to spontaneous SHG. However, there are many possible strategies for alleviating phase instability \cite{opphase1,opphase2,opphase_3} and an improved apparatus is under development.

\begin{figure}[t!]
\centering
\includegraphics[width=.8\columnwidth]{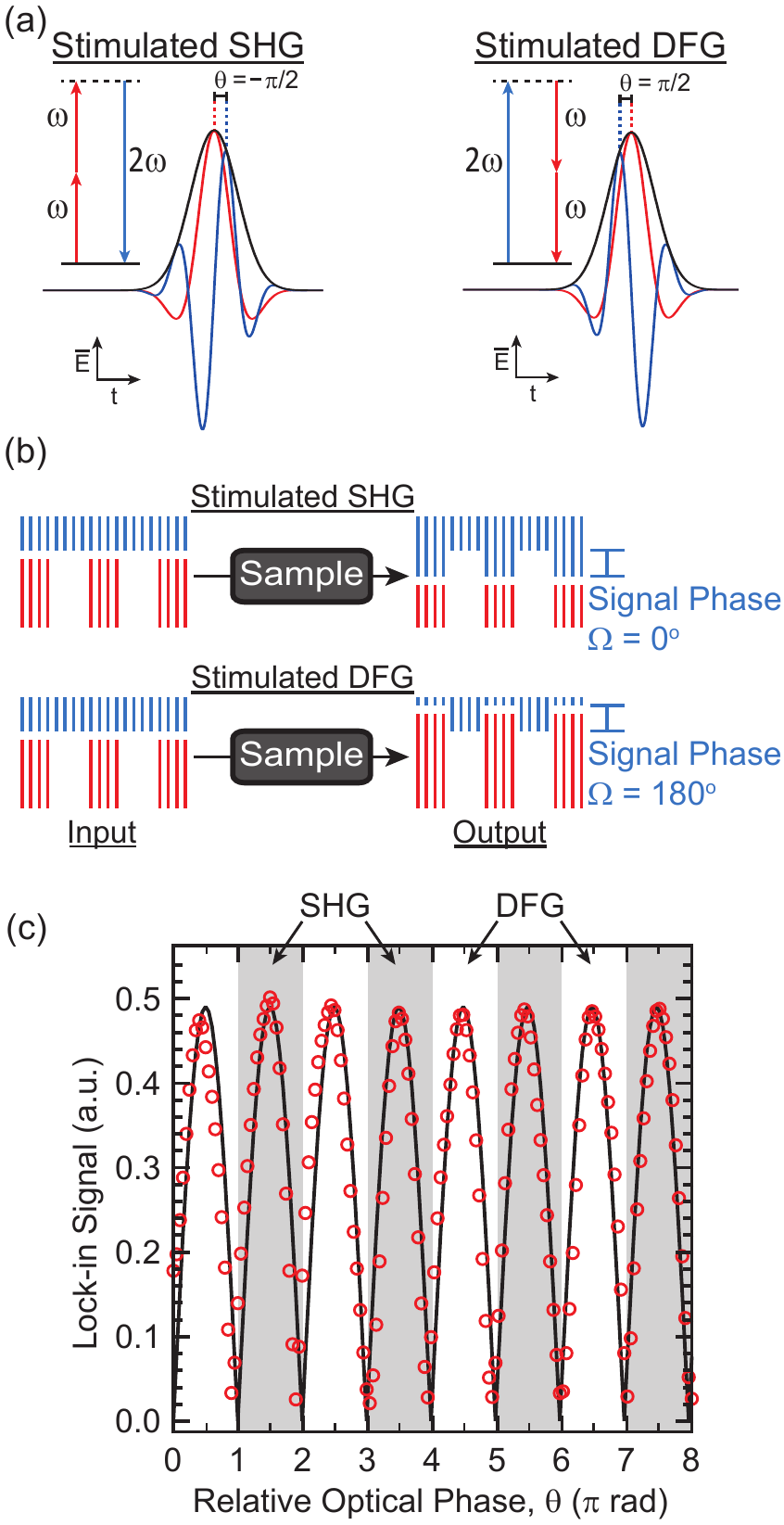}
\caption{(Color online) Effect of phase on stimulated SHG (a) Relative optical phase relations that lead to stimulated SHG (left) and DFG (right). (b) Effect of stimulated SHG and DFG on the signal phase $\Omega$ measured by the lock-in amplifier. (c) Dependence of the lock-in detected signal on the relative optical phase $\theta$. Grey regions indicate an SHG signal in phase with the reference $\left(\Omega=0^\circ\right)$, whereas the white regions represent a stimulated DFG signal measured to be out of phase with the reference $\left(\Omega=180^\circ\right)$.}
\end{figure}

To demonstrate the technique's utility, stimulated SHG signals were collected from bovine collagen I. Collagen I is a naturally abundant nonlinear material that is frequently the target of bioimaging studies. SHG imaging of collagen fibrils can distinguish diseased and wild-type tissue morphologies \cite{Collagen_disease} and has recently been used to determine single fibril diameters smaller than the Abbe limit \cite{Collagen_fibril_diameter_NComm2014}. 

Spontaneous and stimulated SHG signals from collagen I were recorded as a function of incident fundamental power. The results are plotted in Fig.~4. The spontaneous signal grows quadratically with incident power, while the stimulated signal increases linearly. Even at very small incident stimulating fluences (2.7 nJ/cm$^2$), the signal amplification exceeds four orders of magnitude. In all other measurable ways, stimulated SHG in collagen I behaved identically to stimulated SHG in BBO.

It is important to note, however, that in thick, semi-ordered collagen samples, dispersion and additional momentum contributions from spatial frequencies present in the sample morphology would prevent phase matching and complicate analysis \cite{Campagnola_PhaseMatch_Tissue}. Our prepared collagen sample is 1-2 fibrils thick, which is much shorter than the length over which the fundamental and second harmonic fields' relative optical phase would change due to momentum mismatch. Such quasi-perfect phase matching is also realized in surface and interfacial spectroscopy \cite{SFG_Spec_polymer,SFG_Spec_AqueousEnv_ChemRev2002,Tisdale2010_SciencePaper,bad_SHG_Data,Visualize_ChargeSep_OrgBulkHet_ncomm}, where we anticipate stimulated SHG will significantly advance our ability to explore interfacial physics at femtosecond timescales and sub-micron length scales. 

\begin{figure}[t!]
\centering
\includegraphics[width=.7\columnwidth]{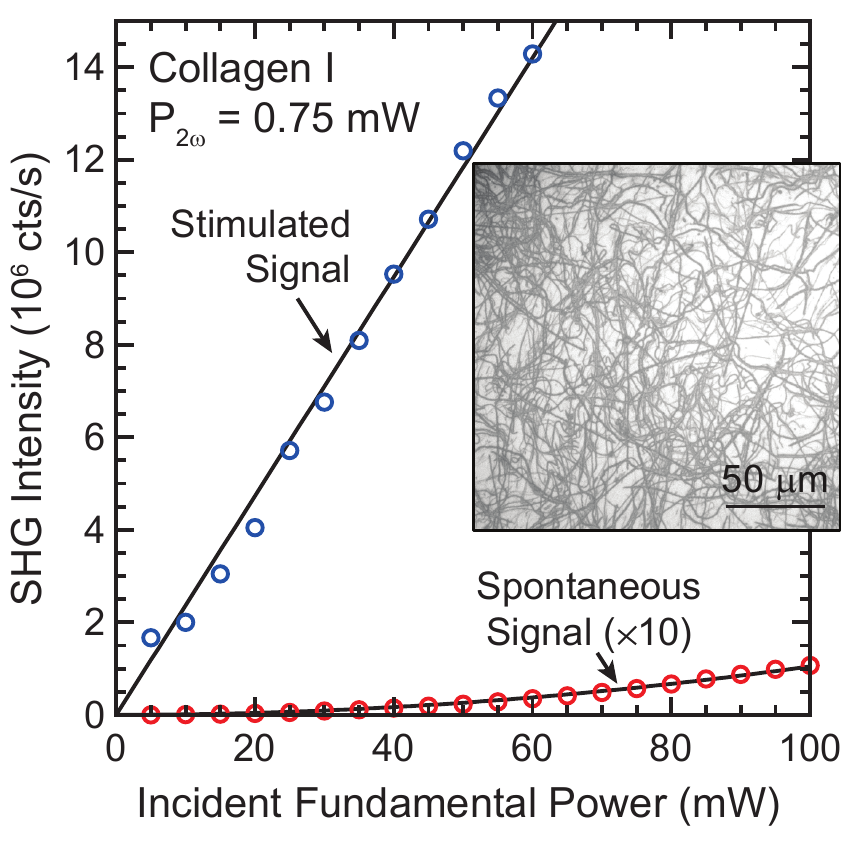}
\caption{(Color online) Illustration of stimulated SHG in collagen I. At low incident powers, stimulated SHG produces more than $10^4$ times as many signal photons as spontaneous SHG. An optical micrograph of the collagen I sample is presented as an inset.}
\end{figure}

Finally, it is worth noting similarities and differences between stimulated SHG and heterodyne SHG \cite{XYZ_heterodyne_SHG,Ogilvie_Heterodyne_OptLett2014,So_HeterodyneImag_OptExp2004,FreqDomainInterferSHSpec_OptLett1999,Downer_2003_JOSAB,phaseAndAmp_nonlinSusc_PRB2007,Phase_sens_NonLin_PRB1998} or SFG \cite{2D_Heterodyne_IRSpec,Heterodyne_SFG_SNR_2008JACS}. Both approaches utilize an additional optical field to amplify an otherwise weak nonlinear optical signal, leading to linear (instead of quadratic) dependence on laser power and interaction volume. Whereas heterodyne SHG typically involves mixing of a time-delayed reference signal with the SHG signal generated from the sample in a spectrometer, stimulated SHG involves mixing of temporally coincident fields within the sample itself. The salient difference is that in stimulated SHG power is actually transferred between two freely propagating fields. This has the notable benefit of allowing one to directly detect the intensity of the SHG field using a single channel detector, such as a photodiode, rendering the method suitable for high-speed imaging. Another potential benefit of the optical stimulation approach is the ability to detect changes in $I_\omega$ instead of $I_{2\omega}$, which could enable access to structures and interfaces buried within media that are absorptive or highly scattering at the second harmonic frequency.

We thank Pooja Tyagi for many helpful discussions. This work was supported by the U.S. Department of Energy, Office of Basic Energy Sciences, under Award Number DE-SC0010538. A.J.G. acknowledges partial support from a National Science Foundation Graduate Research Fellowship under Grant No. 1122374.

\FloatBarrier
\bibliographystyle{apsrev4-1}
\bibliography{Library}

\end{document}